\documentclass[12pt]{article}
\usepackage{amsmath,amssymb,amstext,amsthm,exscale,latexsym}
\input epsf
\newcommand {\al}   {\alpha}       \newcommand {\bt}  {\beta}
       
\newcommand {\dl}   {\delta}       \newcommand {\e }  {\epsilon}
        
\newcommand {\ve}   {\varepsilon}

\newcommand {\s }   {\sigma}       \newcommand {\vr } {\varrho}
         
\newcommand {\vf }  {\varphi}      
         \newcommand {\om}  {\omega}

\newcommand {\pl}   {\partial}     
     
     \newcommand   {\tr}{{\sf\,tr\,}}

\renewcommand {\sin}{{\sf\,sin\,}}       \renewcommand {\cos}{{\sf\,cos\,}}
\newcommand {\tg}{{\sf\,tg\,}}           
 \renewcommand {\arccos}{{\sf\,arccos\,}}

   \newcommand {\MB}  {{\mathbb B}}

\newcommand {\MI }  {{\mathbb I}}   
   
\newcommand {\MM }  {{\mathbb M}}   
\newcommand {\MO }  {{\mathbb O}}   \newcommand {\MP}  {{\mathbb P}}
   \newcommand {\MR}  {{\mathbb R}}
\newcommand {\MS }  {{\mathbb S}}   
\newcommand {\MU }  {{\mathbb U}}   
   
   \newcommand {\MZ}  {{\mathbb Z}}

\newcommand {\Go}  {\mathfrak{o}}   
   
\newcommand {\Gs}  {\mathfrak{s}}   
\newcommand {\Gu}  {\mathfrak{u}}

\newcommand {\Sa}  {{\textsc{a}}}   \newcommand {\Sb}  {{\textsc{b}}}
\newcommand {\CC }  {{\cal C}}      

\begin{document}
\title     {One-dimensional topologically nontrivial \\
            solutions in the Skyrme model}
\author    {M. O. Katanaev
            \thanks{E-mail: katanaev@mi.ras.ru}\\ \\
            \sl Steklov Mathematical Institute,\\
            \sl Gubkin St.~8, 119991, Moscow, Russia}
\date      {15 April 2003}
\maketitle
\begin{abstract}
We consider the Skyrme model using the explicit parameterization
of the rotation group $\MS\MO(3)$ through elements of its algebra.
Topologically nontrivial solutions already arise in the one-dimensional
case because the fundamental group of $\MS\MO(3)$ is $\MZ_2$.
We explicitly find and analyze one-dimensional static solutions.
Among them, there are topologically nontrivial solutions with finite
energy. We propose a new class of projective models whose
target spaces are arbitrary real projective spaces $\MR\MP^d$.
\end{abstract}
\section{Introduction}
The Skyrme model \cite{Skyrme61} is one of the fundamental contemporary
mathematical physics models, and it finds applications in different
fields of physics ranging from the theory of elementary particles to solid
state physics. The model was originally proposed in terms of four
scalar fields taking values on the three-dimensional sphere $\MS^3$.
The Lagrangian of the model was also written in terms of elements
of the algebra $\Gs\Gu(2)$ because the sphere $\MS^3$ is diffeomorphic
to the group $\MS\MU(2)$ as a manifold. Because the fundamental and
the second homotopy groups of unitary groups $\MS\MU(n)$, $n\ge2$, and
spheres $\MS^n$, $n\ge3$, are trivial, one- and two-dimensional
topologically nontrivial solutions of the corresponding models do not exist.
In such a parameterization of the Skyrme model, topologically nontrivial
solutions are related to the third homotopy group $\pi_3(\MS^3)=\MZ$. But
these solutions have not yet been found in explicit form.

In subsequent years, much attention was also given to $\MS\MO(n)$ models
where target spaces are any dimensional spheres $\MS^{n-1}$ with the
rotation groups $\MS\MO(n)$ being the symmetry groups. Reviews of these
models and references can be found in [2--6].
\nocite{ZaMaNoPi80E,Rajara82,TakFad86E,Zakrze89,RybSan01E}%
In what follows, we call these models $\MS^{n}$ models preserving the
term $\MS\MO(n)$ model for the case where the target space is the group
manifold itself.

Topological solitons of the lowest dimension exist for $\MS^1$ models for
which $\pi(\MS^1)=\MZ$. For example, there are well known kinks in
the sine--Gordon model. Topologically nontrivial solutions for higher
homotopy groups appear in the $\MS^2$ model, where the second homotopy
group is nontrivial, $\pi_2(\MS^2)=\MZ$. The corresponding static solutions with
nontrivial topological charge were found and analyzed in \cite{BelPol75}.

As already noted, the Lagrangian for the Skyrme model can be written
in terms of the $\Gs\Gu(2)$ algebra elements. Because the algebras for
the $\MS\MU(2)$ and $\MS\MO(3)$ groups are isomorphic, the Skyrme model
can be considered the $\MS\MO(3)$ model with the group manifold
itself being the target space. For this, we use the known (see, i.g.,
\cite{EllDaw79}), but not widely used, explicit parameterization of the
rotation group $\MS\MO(3)$, i.e., we work with the three-dimensional group
manifold directly. Topologically nontrivial solutions appear here,
even in the one-dimensional case, because the fundamental group of the
rotation group $\MS\MO(3)$ is nontrivial $(\pi(\MS\MO(3))=\MZ_2)$. We
explicitly find the corresponding static solutions. They are analyzed and
compared with the static solutions of the $\MS^2$ model.

The explicit parameterization of the group $\MS\MO(3)$ used in this
paper allows a generalization. It is well known that the
$\MS\MO(3)$ group manifold is diffeomorphic to the three-dimensional
projective space $\MR\MP^3$ (see, e.g., \cite{DuNoFo98E}).
This allows generalizing the Skyrme model to the case of arbitrary
projective spaces over the field of real numbers where the target space is
an arbitrary projective space $\MR\MP^n=\MS^n/\MZ_2$. The projective
space $\MR\MP^n$ can be parameterized by points in the Euclidean space
$\lbrace\om^i\rbrace\in\MR^n$, $i=1,\dotsc,n$, inside the ball
$\om=\sqrt{\om^i\om_i}\le\pi$ for which the antipodal points of the
boundary sphere of radius $\pi$ are identified. In this parameterization
of the projective space, the Lagrangian for an $\MR\MP^n$ model depends on
$n$ fields $\om^i(x)$, and is invariant under local discrete transformations
$\om^i\rightarrow\om^i+2\pi\om^i/\om$. In contrast to gauge models, the
symmetry group at each point of a space-time is a discreet group $\MZ$ of
translations on a constant vector, not a Lie group. Representation of the
translation group is local and depends on the space-time point because
the direction of a vector changes continuously.

We start our consideration by describing the explicit parameterization
of the group $\MO(3)$. In Secs.\ 3, 4, and 5 we write the respective
Lagrangian for $\MS\MO(3)$, $\MS\MU(2)$, and $\MS^2$ models, and we
compare them. Static solutions for the $\MS^2$ and $\MS\MO(3)$ models
are found and compared in Secs.\ 6 and 7. The $\MS\MO(3)$ model is
generalized to arbitrary real projective spaces $\MR\MP^d$ in Sec.~8.
\section{Parameterization of three-dimensional \\ rotation group}
Calculations for the group $\MO(3)$ can be conveniently performed in
the explicit parameterization of group elements by elements of its
algebra. An element of the algebra $\Gs\Go(3)$ can be parameterized
by an arbitrary antisymmetric $3\times3$ matrix
\begin{equation}                                        \label{epeans}
  (\om\ve)_i{}^j=(\om^k \ve_k)_i{}^j=\om^k\ve_{ki}{}^j~~\in\Gs\Go(3),
\end{equation}
where $\ve_{ijk}$ is the totally antisymmetric third rank tensor,
$\ve_{123}=1$, and raising and lowering of indices is performed with
the Kronecker symbol. Here, the first index $k$ enumerates the basis
of the algebra, and the indices $i$ and $j$ are considered matrix indices.
An algebra element is parameterized by a three-dimensional vector
$\om^k\in\MR^3$, and the $\MO(3)$ group is therefore three-dimensional.
There is a single element of the group $\MS\MO(3)$ (from the component
connected to the unity element) that corresponds to an element of the algebra:
\begin{equation}                                        \label{elsogr}
  S_{(+)i}{}^j=(e^{(\om\ve)})_i{}^j=\dl_i^j\cos\om\
  +\frac{(\om\ve)_i{}^j}\om\sin\om
  +\frac{\om_i\om^j}{\om^2}(1-\cos\om)~          \in\MS\MO(3),
\end{equation}
where $\om=\sqrt{\om^i\om_i}$ is the modulus of the vector $\om^i$.
Direct calculations show that $S_{(+)i}{}^j$ is indeed the orthogonal matrix.
Vice versa, any orthogonal matrix with a positive determinant can be
represented in form (\ref{elsogr}) for a vector $\om^i$. We note that
in contrast to the algebra element, the group element has both symmetric
and antisymmetric parts. It can be verified that the $\MS\MO(3)$-group
element is invariant with respect to translation of a vector
$\om^i\rightarrow\om^i+2\pi\om^i/\om$,
\begin{equation*}
  S_{(+)i}{}^j\left(\om^k+2\pi\frac{\om^k}\om\right)=S_{(+)i}{}^j(\om^k),
\end{equation*}
and this is the only invariance. The shift of the vector $\om^i$ changes only
its length $\om\rightarrow\om+2\pi$, leaving its direction unchanged. It
is easy to observe the symmetry of the rotation matrix by noting that the
ratio $\om^i/\om$ defining the direction of vector $\om^i$ remains
unchanged under an arbitrary shift,
\begin{equation*}
  \frac{\om^i}\om\rightarrow\frac{\om^i+c\om^i/\om}{\om+c}=\frac{\om^i}\om,
  ~~~~c\in\MR.
\end{equation*}
An element of the rotation group is therefore parameterized by a point
of the Euclidean space $\om^i\in\MR^3$ with the only equivalence relation
\begin{equation}                                        \label{eqrsog}
  \om^i\approx\om^i+2\pi\frac{\om^i}\om.
\end{equation}
Here, the uncertainty at zero is resolved along radial directions
$\om^i=\e k^i$, $\e\to0$. The origin of the coordinate system
is thus identified with the spheres of radii $2\pi m$, $m=1,2,\dotsc$.

The vector $\om^i$ parameterizes the $\MS\MO(3)$ as follows. The direction
of $\om^i$ coincides with the rotation axis, and its modulus $\om$ equals
the rotation angle. Each $\MS\MO(3)$-group element is then identified with a
point of the three-dimensional ball $\MB^3_\pi$ having the radius $\pi$ and
the center at the origin. Different rotations correspond to different
points, and the antipodal points of the boundary sphere must be identified
because rotations through the angles $\pi$ and $-\pi$ about a fixed axis
yield the same result.

The full group of three-dimensional rotations $\MO(3)$ consists of two
disconnected components: orthogonal matrices with positive $S_{(+)}$ and
negative $S_{(-)}$ determinants. Elements of the full group $\MO(3)$ are
parameterized, for example, by algebra elements (\ref{epeans}) as
\begin{equation}                                        \label{elsogt}
  S_{(\pm)i}{}^j=\dl_i^j\cos\om+\frac{(\om\ve)_i{}^j}\om\sin\om
  +\frac{\om_i\om^j}{\om^2}(\pm1-\cos\om)~\in\MO(3),
\end{equation}
We note that there are two group elements $S_{(\pm)}\in\MO(3)$,
one from each component, which correspond to an element of the
algebra $(\om\ve)_i{}^j\in\Gs\Go(3)$.

The inverse matrices have the form
\begin{equation}                                        \label{elsogi}
  S^{-1}_{(\pm)i}{}^j(\om^k)=S_{(\pm)i}{}^j(-\om^k)=\dl_i^j\cos\om
  -\frac{(\om\ve)_i{}^j}\om\sin\om
  +\frac{\om_i\om^j}{\om^2}(\pm1-\cos\om)~\in\MO(3),
\end{equation}
i.e., they correspond to the inverse vector $-\om^i\in\MR^3$. In
other words, the inverse group element corresponds to the rotation of
the Euclidean space about the same axis through the opposite angle.

Contraction of the matrix $S_{(\pm)i}{}^j$ with the vector $\om_j$ yields
\begin{equation*}
  S_{(\pm)i}{}^j\om_j=\pm\om_i.
\end{equation*}
This means that the vector $\om^i$ is the eigenvector of orthogonal matrices
with the eigenvalues $\pm1$. In other words, rotations $S_{(+)}$ leave the
direction of the rotation axis unchanged, while rotations $S_{(-)}$ change
it to the opposite direction.

We assume that $\MO(3)$-group element depends smoothly on a point of an
arbitrary manifold $\MM$, i.e.,
\begin{equation*}
  \om^i(x):~~\MM\rightarrow\MO(3).
\end{equation*}
We introduce the notation
\begin{equation}                                        \label{ealroe}
  l_{(\pm)\al i}{}^j=(\pl_\al S_{(\pm)}^{-1}S_{(\pm)})_i{}^j.
\end{equation}
The validity of the formula
\begin{equation}                                           \label{ealgrs}
\begin{split}
  l_{(\pm)\al i}{}^j=\mp\frac{(\pl_\al\om\ve)_i{}^j}\om\sin\om
  -\frac{\pl_\al\om(\om\ve)_i{}^j}\om\left(1\mp\frac{\sin\om}\om\right)&
\\
  \pm\frac{\pl_\al\om_i\om^j-\om_i\pl_\al\om^j}{\om^2}(\pm1-\cos\om)&
  ~~\in\Gs\Go(3)
\end{split}
\end{equation}
can be shown by straightforward calculations. This matrix is antisymmetric
with respect to its indices and is hence an element of the algebra $\Gs\Go(3)$.
It is the trivial $\MO(3)$ connection (pure gauge) for which the curvature
tensor is zero. The trivial $\MS\MO(3)$ connection is denoted by
$l_{\al i}{}^j=l_{(+)\al i}{}^j$ in what follows.

The considered parameterization of the rotation group shows that the
group manifold is compact and orientable. The component $\MS\MO(3)$
connected to the unity element, as a manifold, is diffeomorphic to the
three-dimensional projective space $\MR\MP^3$. It is not simply
connected, and its fundamental group is $\MZ_2$.
\section{The Lagrangian for the Skyrme model}
We consider the $n$-dimensional Minkowski space $\MR^{1,n-1}$ with
Cartesian coordinates $x^\al$, $\al=0,1,\dotsc,n-1$. Let an element
of the rotation group $\om^i(x)$ be given at each point of the Minkowski
space. To describe $\pi$-mesons Skyrme proposed the action \cite{Skyrme61}
\begin{equation}                                        \label{eskact}
  S=\int dx L=\int dx\left(-\frac\vr4\tr(l_\al l^\al)
  -\frac\kappa8\tr([l_\al,l_\bt][l^\al,l^\bt])\right),
\end{equation}
where $\vr$ and $\kappa$ are coupling constants and square brackets
denote the matrix commutator. Skyrme considered $l_\al$ as a trivial
$\MS\MU(2)$ connection and parameterized it by a four-vector in the
isotopic space with the four-vector taking values on a three-dimensional
sphere $\MS^3$, i.e., he considered action (\ref{eskact}) as the
$\MS\MU(2)$ model or, equivalently, as the $\MS^3$ model. We consider
(\ref{eskact}) as the action for the $\MS\MO(3)$ model using the
parameterization of the rotation group from the previous section.
We note that these models are not equivalent. Substituting the explicit
expression for $l_\al$ in terms of the group parameters (\ref{ealgrs})
yields the Lagrangian for three independent scalar fields $\om^i$,
\begin{equation}                                        \label{eacomi}
\begin{split}
  L&=\vr\left(\frac12\pl_\al\om^2
  +(1-\cos\om)\frac{(\pl_\al\om\pl_\al\om)-\pl_\al\om^2}{\om^2}\right)
\\
  &+\kappa\left((1-\cos\om)\frac{(\pl_\al\om\pl_\al\om)\pl_\bt\om^2
  -(\pl_\al\om\pl_\bt\om)\pl_\al\om\pl_\bt\om}{\om^2}\right.
\\
  &+\left.(1-\cos\om)^2\frac{(\pl_\al\om\pl_\al\om)^2
  -(\pl_\al\om\pl_\bt\om)^2
  -2(\pl_\al\om\pl_\al\om)\pl_\bt\om^2
  +2(\pl_\al\om\pl_\bt\om)\pl_\al\om\pl_\bt\om}{\om^4}\right),
\end{split}
\end{equation}
where we introduce the notation for the scalar product
\begin{equation*}
  (\pl_\al\om\pl_\al\om)=\pl_\al\om^i\pl_\al\om_i\ne\pl_\al\om^2.
\end{equation*}
In this expression, repeated Greek indices imply summation with the
Minkowski metric tensor$\eta_{\al\bt}$, and we do not distinguish
lower and upper indices for simplicity. The three fields $\om^i(x)$ in
the Lagrangian are independent dynamical variables free of constraints
except equivalence relation (\ref{eqrsog}). These fields are scalars
under coordinate transformations in the Minkowski space $\MR^{1,n-1}$
and vector components under global rotations of the Euclidean
target-space $\MR^3$. It can be verified that Lagrangian (\ref{eacomi})
is invariant under the Poicar\'e group acting in the Minkowski space,
global $\MO(3)$ rotations acting in the target space $\om^i\in\MR^3$,
and local translations (\ref{eqrsog}). The latter transformation implies
that the length of the vector $\om^i\in\MR^3$ at each point
$x\in\MR^{1,n-1}$ can be changed by a constant value divisible by $2\pi$.
At the same time, this discrete transformation is local because the
direction of the vector $\om^i$ changes from point to point. By virtue of
this invariance, we assume a vector $\om^i$ to take values inside the
ball of radius $\pi$ with the center at the origin of the target space
$\om^i\in\MB^3_\pi\subset\MR^3$.

To analyze the Skyrme model in form (\ref{eacomi}), it is convenient to
consider four new variables: the length of the vector $\om^i$ (the rotation
angle $\om$) and its direction (the rotation axis $k^i$), not the vector
$\om^i$ itself
\begin{equation}                                        \label{enewvs}
  \om=\sqrt{(\om^1)^2+(\om^2)^2+(\om^3)^2},~~~~k^i=\frac{\om^i}\om.
\end{equation}
By definition, the vector $k^i$ has unit length $(k,k)=k^ik_i=1$.
In terms of the new variables, Lagrangian (\ref{eacomi}) becomes
\begin{equation}                                        \label{elaomn}
\begin{split}
  L&=\vr\left(\frac12\pl_\al\om^2+(1-\cos\om)(\pl_\al k\pl_\al k)\right)
\\
  &+\kappa\left(\vphantom{\frac12}(1-\cos\om)
  \left[(\pl_\al k\pl_\al k)\pl_\bt\om^2
  -(\pl_\al k\pl_\bt k)\pl_\al\om\pl_\bt\om\right]\right.
\\
  &\left.\vphantom{\frac12}~~~~~~~~~~~~~~~
  +(1-\cos\om)^2\left[(\pl_\al k\pl_\al k)^2
  -(\pl_\al k\pl_\bt k)^2\right]\right).
\end{split}
\end{equation}
We recall that $(\pl_\al k\pl_\al k)=\pl_\al k^i\pl_\al k_i$.
Transformation (\ref{eqrsog}) changes only the length of a vector
\begin{equation}                                       \label{einomn}
  \om\rightarrow\om+2\pi,~~~~k^i\rightarrow k^i.
\end{equation}
The invariance of Lagrangian (\ref{elaomn}) under transformations
(\ref{einomn}) follows immediately. Moreover,
each of its terms is separately invariant.
\section{Comparison with the $\MS\MU(2)$ model}
The form of the action of Skyrme model (\ref{eskact}) for group parameters
depends only on an algebra, and not on the group that represents the
target space. To show the dependence of the model on the group, we
consider the $\MS\MU(2)$ model. An element of the algebra $\Gs\Gu(2)$
has the form
\begin{equation*}
  \om^k\frac i2(\s_k)_\Sa{}^\Sb~\in\Gs\Gu(2),~~~~\Sa,\Sb=1,2,
\end{equation*}
where $\s_k$ are the Pauli matrices. We introduce the factor $i/2$ for
the commutators of the basis vectors for the algebras $\Gs\Gu(2)$ and
$\Gs\Go(3)$ to have the same form. The corresponding $\MS\MU(2)$ group
element is
\begin{equation}                                        \label{esutgr}
  U_\Sa{}^\Sb=(e^{(i\om\s/2)})_\Sa{}^\Sb
  =\dl_\Sa^\Sb\cos\frac\om2+i\frac{\om^k\s_{k\Sa}{}^\Sb}\om\sin\frac\om2
  ~~\in\MS\MU(2).
\end{equation}
From this expression for a group element, we obtain the identity
\begin{equation*}
  U_\Sa{}^\Sb\left(\om^i+4\pi\frac{\om^i}\om\right)=U_\Sa{}^\Sb(\om^i),
\end{equation*}
i.e., there exists the equivalence relation
\begin{equation}                                        \label{eqrsut}
  \om^i\approx \om^i+4\pi\frac{\om^i}\om
\end{equation}
in the parameter space for the group $\MS\MU(2)$. In comparison with
equivalence relation (\ref{eqrsog}) for $\MS\MO(3)$, the shift is performed
by a vector having a double length. This means that the group manifold is
parameterized by internal points of the ball $\MB^3_{2\pi}$ of radius $2\pi$.
Moreover, all points of the boundary sphere must be identified. This is the
second equivalence relation in the parameter space,
\begin{equation}                                       \label{equspe}
  \om_1^i|_{\om_1=2\pi}\approx\om_2^i|_{\om_2=2\pi}.
\end{equation}
The extra equivalence relation is due to the absence of the third term
in (\ref{esutgr}) as compared with (\ref{elsogr}). It is another difference
from the rotation-group case, where only antipodal points on the boundary
sphere are identified, which is taken into account in equivalence relation
(\ref{eqrsog}). It can be shown that there are no other equivalence
relations in the parameter space except (\ref{eqrsut}) and (\ref{equspe}).

It can be easily verified that the Lagrangian for the Skyrme model has the
same form (\ref{elaomn}) for both the groups $\MS\MU(2)$ and $\MS\MO(3)$.
The difference amounts to equivalence relations (\ref{eqrsog}) and
(\ref{eqrsut}) and identifications of points on boundary sphere (\ref{equspe}).
The last property results in the difference in fundamental groups leading to
the existence of one-dimensional topologically nontrivial solutions in
the $\MS\MO(3)$ Skyrme model.
\section{$\MS^2$ model}
The $\MS^2$ model has attracted much interest for many years. Because
both the $\MS\MO(3)$ Skyrme model and $\MS^2$ model are described by
three-dimensional vectors, and any vector of a fixed length can be
represented as a result of action of the rotation matrix on some fixed
vector of the same length, there arises a question about a possible
relation between these models.  In spite of the nonequivalence of these
models, there is a relation between them, which we consider in this
section.

We relate the $\MS\MO(3)$ in Sec.~3 to the $\MS^2$ model. Given a unit
vector $n^i(x)$, $(n,n)=1$, at each point of the Minkowski space $\MR^{1,n-1}$,
we obtain the target space, which is the two-dimensional sphere $\MS^2$.
We let the capital letters $X,Y$, and $Z$ denote the Cartesian coordinates
of the target space. We consider the Lagrangian proposed by Faddeev for
the $\MS^2$ model \cite{Faddee77}
\begin{equation}                                        \label{ectmfi}
\begin{split}
  L_\mathrm{F}&=m^2(\pl_\al n\pl_\al n)
  +\frac1{e^2}(n^i\ve_{ijk}\pl_\al n^j\pl_\bt n^k)^2,
\\
  &=m^2(\pl_\al n\pl_\al n)
  +\frac1{e^2}\left[(\pl_\al n\pl_\al n)^2-(\pl_\al n\pl_\bt n)^2\right].
\end{split}
\end{equation}
This model has topologically nontrivial solutions related to the third
homotopy group $\pi_3(\MS^2)=\MZ$ [11--15].
\nocite{FadNie99A,FadNie97,BatSut98,BatSut99,HieSal99}

Any unit vector $n\in\MS^2$ can be parameterized by an element of the group
of three-dimensional rotations
\begin{equation}                                        \label{emintn}
  n^i=n^j_0S_j{}^i(\om),
\end{equation}
where $n^i_0$ is an arbitrary fixed vector of unit length. This
correspondence is not one-to-one, because two given vectors $n$ and $n_0$
do not define the rotation-group element uniquely. Looking ahead, we
say that the $\MS\MO(3)$ and $\MS^2$ models are not equivalent.
Lagrangian (\ref{ectmfi}) in parameterization (\ref{emintn}) becomes
\begin{equation*}
\begin{split}
  L&=m^2\, n_0^in_0^j\pl_\al S_i{}^k\pl_\al S_{jk}
\\
  &+\frac1{e^2}\,n_0^i n_0^j n_0^k n_0^l\left(
  \pl_\al S_i{}^m\pl_\al S_{jm}\pl_\bt S_k{}^n\pl_\bt S_{ln}
  -\pl_\al S_i{}^m\pl_\bt S_{jm}\pl_\bt S_k{}^n\pl_\al S_{ln}\right).
\end{split}
\end{equation*}
It depends on three scalar fields $\om^i(x)$ parameterizing the rotation
group and on a fixed vector $n_0$. We average it over angles
defining the vector $n_0$. For this, we use the averaging formulas
(which can be verified straightforwardly)
\begin{equation*}
\begin{split}
  \langle n_0^in_0^j\rangle&
  =\frac1{4\pi}\int d\Theta d\Phi\,\sin\Theta n_0^in_0^j=\frac13\dl^{ij},
\\
  \langle n_0^in_0^jn_0^kn_0^l\rangle&
  =\frac1{4\pi}\int d\Theta d\Phi\,\sin\Theta n_0^in_0^jn_0^kn_0^l
  =\frac1{15}(\dl^{ij}\dl^{kl}+\dl^{ik}\dl^{jl}+\dl^{il}\dl^{jk}),
\end{split}
\end{equation*}
where $\Theta,\Phi$ are polar coordinates in the target space. After
averaging, we obtain the new Lagrangian
\begin{equation*}
\begin{split}
  \langle L_\mathrm{F}\rangle=&-\frac{m^2}3\tr(l_\al l_\al)
\\
  &-\frac1{15e^2}\left[\frac12\tr([l_\al,l_\bt][l_\al,l_\bt])
  -\tr^2(l_\al l_\al)+\tr^2(l_\al l_\bt)\right].
\end{split}
\end{equation*}
It contains two additional terms in comparison with Skyrme model
Lagrangian (\ref{eskact}).

Averaging the $\MS^2$ model over the directions of the vector $n_0$
therefore yields the $\MS\MO(3)$ model. The nonequivalence of these
models is apparent for the static solutions considered in the
subsequent sections.
\section{Static solutions for the two-dimensional $\MS^2$ model}
The unit vector $n$ in spherical coordinates is parameterized by two
angles $\Theta$ and $\Phi$,
\begin{equation}                                        \label{epacsh}
  n^1=\sin\Theta\cos\Phi,~~~~n^2=\sin\Theta\sin\Phi,~~~~n^3=\cos\Theta,
\end{equation}
describing two independent degrees of freedom. For simplicity, we
analyze static solutions in the two-dimensional $\MS^2$ model.
In this case, the second term in Lagrangian (\ref{ectmfi}) does
not contribute to the equations for static solutions. Therefore, we
restrict ourselves to considering only the first term.

In angle coordinates, the action corresponding to the first term in
(\ref{ectmfi}) becomes
\begin{equation}                                        \label{ecphth}
  S=m^2\int dx(\pl_\al\Theta^2+\sin^2\Theta\pl_\al\Phi^2)
\end{equation}
and yields the equations of motion
\begin{align}                                           \label{evarth}
  \frac1{m^2}\frac{\dl S}{\dl\Theta}=~~~~
  &-2\square\Theta+2\sin\Theta\cos\Theta\pl_\al\Phi^2=0,
\\                                                      \label{evaphi}
  \frac1{m^2}\frac{\dl S}{\dl\Phi}=~~~~
  &-2\pl_\al(\pl_\al\Phi\sin^2\Theta)=0,
\end{align}
where $\square=\pl_\al\pl_\al$ is the d'Alembert operator.

We consider the two-dimensional Minkowski space $\MR^{1,1}$ with
Cartesian coordinates $t,x$ and analyze static solutions. For
$\Theta=\Theta(x)$ and $\Phi=\Phi(x)$ equations of motion (\ref{evarth})
and (\ref{evaphi}) become
\begin{align}                                           \label{estthe}
  \Theta''-\sin\Theta\cos\Theta\Phi^{\prime 2}&=0,
\\                                                      \label{estphi}
  \left(\Phi'\sin^2\Theta\right)'&=0.
\end{align}
where the prime denotes differentiation with respect to $x$. We seek
solutions of the system of ordinary differential equations in the class
of doubly differentiable functions $\CC^2(L)$ on a finite interval
$x\in[0,L]$. Any solution of this system of equations belongs to one
of the following four classes (the integration constants are
denoted by $b,c,x_0$, and $\Phi_0$ below).

{\bf Degenerate solutions I.}
\begin{equation}                                        \label{efideg}
  \Theta=0,~~~~\Phi(x)\in\CC^2(L).
\end{equation}
In this case, the vector $n$ is directed along the $Z$ axis, and there
is no rotation.

{\bf Degenerate solutions II.}
\begin{equation}                                        \label{esedeg}
  \Theta=\frac\pi2,~~~~\Phi=bx+x_0.
\end{equation}
In this case, the vector $n$ rotates uniformly in the $XY$ plane as it
moves along the $x$ coordinate.

{\bf Degenerate solutions III.}
\begin{equation}                                        \label{ethdeg}
  \Theta=bx+x_0,~~~~\Phi=\Phi_0.
\end{equation}
Here, the vector $n$ rotates uniformly in the plane $\Phi=\Phi_0$ as it moves
along $x$. These solutions essentially coincide with degenerate solutions II.
To show this, we can use the freedom in choosing the coordinate system in the
target space and choose a new $Z'$ axis perpendicular to the plane
$\Phi=\Phi_0$. Degenerate solutions III then become degenerate solutions II.

{\bf A general situation.}
In the case of a general situation, Eq.~(\ref{estphi}) has a solution
\begin{equation}                                       \label{ephpti}
  \Phi'\sin^2\Theta=c,~~~~c\ne0.
\end{equation}
Equation (\ref{estthe}) then becomes
\begin{equation*}
  \Theta''=c^2\frac{\cos\Theta}{\sin^3\Theta}.
\end{equation*}
It can be easily integrated,
\begin{equation}                                       \label{esolth}
  \cos\Theta=\sqrt{1-\frac{c^2}{b^2}}\sin[b(x+x_0)],~~~~|b|>|c|.
\end{equation}
We then integrate Eq.~(\ref{ephpti}) and obtain
\begin{equation}                                        \label{esophi}
  \tg(\Phi+\Phi_0)=\frac cb\tg[b(x+x_0)].
\end{equation}
Formulas (\ref{esolth}) and (\ref{esophi}) therefore yield static solutions
in a general situation. The integration constants $x_0$ and $\Phi_0$
correspond to coordinate shifts and can be set to zero for simplicity.
The constant $b$ corresponds to choosing of the scale of the $x$ coordinate.
For a fixed length $L$ of the interval, it cannot be changed. We note
that degenerate solutions II can be obtained from the general situation
in the limit $|b|=|c|$.

It can be shown that in a general situation, the vector $n$ rotates
uniformly in the plane tilted at the angle $\arccos(c/b)$ to the $XY$
plane as this vector moves along $x$. This means that we in fact have
only two essentially different solutions of the Euler--Lagrange equations.
First, there are degenerate solutions I, where the vector $n$ does not
change when moving along $x$. Second, all other solutions describe uniform
rotation of the vector $n$ in an arbitrarily, but fixed, plane. In this case,
the rotation axis is always perpendicular to the vector $n$. Segregating
these solutions into degenerate and general solutions depends upon the
orientation of the coordinate system with respect to the plane of rotation.
We note that in all cases, the orientation of the Cartesian coordinates
axes $X,Y$, and $Z$ in the target space can be chosen arbitrary with
respect to the space coordinate $x$.

All obtained solutions to the $\MS^2$ model are homotopic to zero because
the fundamental group is trivial, $\pi(\MS^2)=0$. This means that there are
no one-dimensional topologically nontrivial solutions in $\MS^2$ models.

The energy of the two-dimensional $\MS^2$ model for static solutions is
\begin{equation*}
  E=\int_0^L dx\left(\Theta^{\prime2}+\sin^2\Theta\Phi^{\prime2}\right).
\end{equation*}
A straightforward substitution of solutions into this integral yields the
values
\begin{equation*}
  E=\left\lbrace
\begin{aligned}
  0, &~~~~\text{degenerate solutions I},
\\
  b^2L,&~~~~\text{all other solutions}.
\end{aligned} \right.
\end{equation*}
The energy of static solutions is therefore finite for finite $L$. For
$bL=2\pi m$, $m=1,2,\dotsc$, the solutions are periodic. In this case, we
can be consider them as given on a circle.

We analyze the stability of the obtained solutions. The second
variation of the energy has the form
\begin{equation*}
  \dl^2 E=\int_0^Ldx\left(\dl\Theta^{\prime2}
  +\cos(2\Theta)\Phi^{\prime2}\dl\Theta^2
  +2\sin(2\Theta)\Phi'\dl\Theta\dl\Phi'
  +\sin^2\Theta\dl\Phi^{\prime2}\right).
\end{equation*}
Here, the variations $\dl\Theta,\dl\Theta',\dl\Phi$, and $\dl\Phi'$ must be
considered independent functions. A simple analysis shows that
degenerate solutions I yield the absolute minimum energy, and the
remaining solutions correspond to saddle points.
\section{Static solutions of the two-dimensional $\MS\MO(3)$ \\ Skyrme model}
The Euler--Lagrange equations for $\MS\MO(3)$ Skyrme model (\ref{elaomn})
have the forms
\begin{align}                                               \nonumber
  &\frac{\dl S}{\dl\om}=
\\                                                          \nonumber
  &-\vr\square\om+\vr\sin\om(\pl_\al k\pl_\al k)
\\                                                          \nonumber
  &-2\kappa(1-\cos\om)\left[(\pl_{\al\bt}^2k\pl_\al k)\pl_\bt\om
  +(\pl_\al k\pl_\al k)\square\om-(\pl_\al k\square k)\pl_\al\om
  -(\pl_\al k\pl_\bt k)\pl_\al\om\pl_\bt\om\right]
\\                                                          \nonumber
  &-\kappa\sin\om\left[(\pl_\al k\pl_\al k)\pl_\bt\om^2
  -(\pl_\al k\pl_\bt k)\pl_\al\om\pl_\bt\om\right]
\\                                                      \label{eskome}
  &+2\kappa(1-\cos\om)\sin\om\left[(\pl_\al k\pl_\al k)^2
  -(\pl_\al k\pl_\bt k)^2\right]=0,
\\                                                         \nonumber
  &\frac{\dl S}{\dl k_i}=
\\                                                         \nonumber
  &-2\vr(1-\cos\om)\square k^i
  -2\vr\sin\om\pl_\al k^i\pl_\al\om
  -2\vr(1-\cos\om)(\pl_\al k\pl_\al k)k^i
\\                                                         \nonumber
  &-2\kappa(1-\cos\om)\left[\square k^i\pl_\al\om^2
  +\pl_\al k^i\pl_{\al\bt}^2\om\pl_\bt\om
  -\pl_{\al\bt}^2 k^i\pl_\al\om\pl_\bt\om
  -\pl_\al k^i\square\om\pl_\al\om\right]
\\                                                        \nonumber
  &-4\kappa(1-\cos\om)^2\left[(\pl_\al k\pl_\al k)\square k^i
  +(\pl_{\al\bt}^2k\pl_\bt k)\pl_\al k^i
  -(\square k\pl_\al k)\pl_\al k^i
  -(\pl_\al k\pl_\bt k)\pl_{\al\bt}^2k^i\right]
\\                                                        \nonumber
  &-8\kappa(1-\cos\om)\sin\om\left[(\pl_\bt k\pl_\bt k)\pl_\al k^i
  -(\pl_\al k\pl_\bt k)\pl_\bt k^i\right]\pl_\al\om
\\                                                        \nonumber
  &-2\kappa(1-\cos\om)\left[(\pl_\al k\pl_\al k)\pl_\bt\om^2
  -(\pl_\al k\pl_\bt k)\pl_\al\om\pl_\bt\om\right]k^i
\\                                                    \label{esenie}
  &-4\kappa(1-\cos\om)^2
  \left[(\pl_\al k\pl_\al k)^2-(\pl_\al k\pl_\bt k)^2\right]k^i=0.
\end{align}
When varying the action with respect to $k_i$, we use the formula
(which takes the constraint $(k,k)=1$ into account)
\begin{equation*}
  \frac{\dl S}{\dl k_i}=\frac{\bar\dl S}{\bar\dl k_j}(\dl^i_j-k^ik_j),
\end{equation*}
where the variational derivative in the right hand side must be taken
without assuming the constraint.

It can be easily verified that in the case of static solutions $\om(x)$,
$k^i(x)$ in two-dimensional Minkowski space-time $\MR^{1,1}$, all terms
depending on the coupling constant $\kappa$ cancel. This means that the
analysis of one-dimensional static solutions in the $\MS\MO(3)$ model
reduces to studying the Lagrangian
\begin{equation}                                        \label{elalth}
  L=\vr\left[\frac12\pl_\al\om^2
  +(1-\cos\om)\left(\pl_\al\theta^2+\sin^2\theta\pl_\al\vf^2\right)\right],
\end{equation}
where we use the spherical coordinates in the target space
\begin{equation}                                        \label{epacsn}
  k^1=\sin\theta\cos\vf,~~~~k^2=\sin\theta\sin\vf,~~~~k^3=\cos\theta.
\end{equation}
We use lower-case Greek letters for angle coordinates of the unit vector
$k$ to distinguish it from the vector $n$ considered in the previous section.

The equations of motion in the static case are
\begin{align}                                           \label{eqomeg}
  \frac1\vr\frac{\dl S}{\dl\om}=~~~~
  &-\om''+\sin\om(\theta^{\prime2}+\sin^2\theta\vf^{\prime2})=0,
\\                                                      \label{eqthet}
  \frac1\vr\frac{\dl S}{\dl\theta}=~~~~
  &-2\left((1-\cos\om)\theta'\right)'
  +2(1-\cos\om)\sin\theta\cos\theta\vf^{\prime2}=0,
\\                                                      \label{eqphil}
  \frac1\vr\frac{\dl S}{\dl\vf}=~~~~
  &-2\left((1-\cos\om)\sin^2\theta\vf'\right)'=0.
\end{align}
We seek solutions of this system of differential equations in the
function space $\CC^2(L)$. Any solution of system of equations
(\ref{eqomeg})--(\ref{eqphil}) belongs to one of the following five classes
(the integration constants are denoted by $a,b,c,x_0,\vf_0,\theta_0,$
and $u_0$ below).

{\bf Degenerate solutions I}.
In this case, the rotation angle is a linear function
\begin{equation*}
  \om=bx+x_0,
\end{equation*}
and the rotation axis is fixed,
\begin{equation*}
  \theta=\theta_0,~~\vf=\vf_0,~~~~\text{or}~~~~\theta=0,~~\vf(x)\in\CC^2(L).
\end{equation*}
For this solution, the vector $n^i=n_0^jS_j{}^i(\om)$ rotates uniformly
about arbitrary directed vector $k$ as it moves along $x$. In contrast to
the $\MS^2$ model, the vector $n$ is not necessarily perpendicular
to the rotation axis.

\begin{figure}[htb]
 \begin{center} \leavevmode \epsfysize=40mm
 \epsfbox{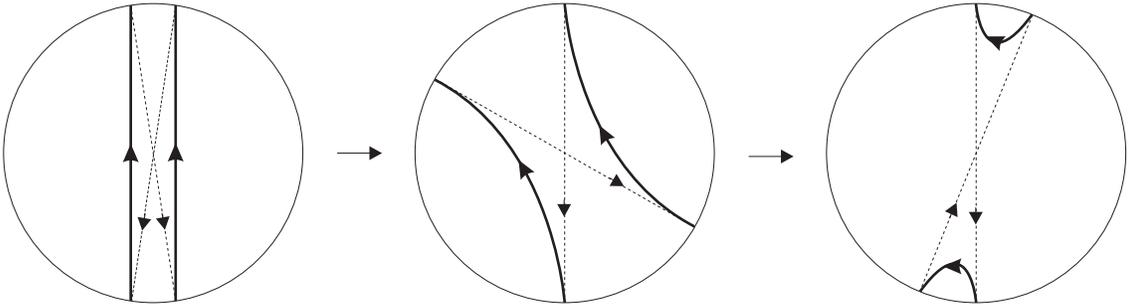}
 \end{center}
 \caption{The continuous deformation of the contour $0<\om<4\pi$ to a point.
 The deformation is shown in the plane containing the vector $k$ determining
 the rotation axis. A dashed line denotes the identification of antipodal
 points of the boundary circle.
 \label{fsohom}}
\end{figure}

Degenerate solutions I yield an example of static topologically nontrivial
solutions with topological charge 0 or 1. If the angle $\om$ varies within
an interval divisible by $4\pi$, then the topological charge is zero because
the corresponding contour can be always continuously deformed to a point.
For $0<\om<4\pi$, the related continuous deformation of the contour is shown
in Fig.~\ref{fsohom} in the plane containing the vector $k$ determining the
rotation axis. A dashed line denotes the identification of antipodal points
of the boundary circle.
If the angle $\om$ takes values in an interval containing the interval
$2\pi$ an odd number of times, then the topological charge equals 1.

{\bf Degenerate solutions II}.
\begin{equation}                                        \label{ethisk}
  \om=0,~~\theta(x),\vf(x)\in\CC^2(L).
\end{equation}
Rotation is absent in this series of solutions because the rotation
angle equals zero.

{\bf Degenerate solutions III}.
\begin{equation*}
\begin{split}
  \cos\frac\om2&=\sqrt{1-\frac{c^2}{4b^2}}\sin[b(x+x_0)],~~~~2|b|\ge|c|\ne0,
\\
  \theta&=\frac\pi2,
\\
  \tg(\vf+\vf_0)&=\frac c{2b}\tg[b(x+x_0)].
\end{split}
\end{equation*}
The polar angle $\vf$ is a monotonic function of $x$ for these solutions,
i.e., the rotation axis $k^i$ rotates in the $XY$ plane. Here,
the rotation angle changes within the finite interval
\begin{equation*}
  2\arccos\sqrt{1-\frac{c^2}{4b^2}}<\om
  <2\pi-2\arccos\sqrt{1-\frac{c^2}{4b^2}}.
\end{equation*}
It can be easily verified that these solutions are homotopic to zero.

{\bf Degenerate solutions IV}.
\begin{equation*}
\begin{split}
  \cos\frac\om2&=\sqrt{1-\frac{c^2}{4b^2}}\sin[b(x+x_0)],~~~~2|b|\ge|c|\ne0,
\\
  \tg(\theta+\theta_0)&=\frac c{2b}\tg[b(x+x_0)],
\\
  \vf&=\vf_0.
\end{split}
\end{equation*}
These solutions essentially coincide with degenerate solutions III,
but the rotation axis rotates in the $\vf=\vf_0$ plane.

{\bf A general situation}.
To solve system of equations (\ref{eqomeg})--(\ref{eqphil}) in a
general situation, we introduce the new coordinate
\begin{equation}                                      \label{eintux}
  u=\int^x\frac{dy}{1-\cos\om(y)}.
\end{equation}
The coordinate $u$ increases with increasing of $x$ because
$dx/du=1-\cos\om\ge0$. Equations (\ref{eqthet}) and (\ref{eqphil}) then
coincide exactly with Eqs.~(\ref{estthe}) and (\ref{estphi}) for
$\Theta$ and $\Phi$, in which differentiation with respect to $x$
must be replaced with the differentiation with respect to $u$. Because
these equations were analyzed in the previous section, it suffices to
solve only Eq.~(\ref{eqomeg}). In a general situation, it becomes
\begin{equation}                                        \label{eqomca}
  \om''=c^2\frac{\sin\om}{(1-\cos\om)^2},~~~~c\ne0.
\end{equation}
A general solution of this equation is
\begin{equation}                                        \label{eqomxc}
  \cos\frac\om2=\sqrt{1-\frac{c^2}{4b^2}}\sin[b(x+x_0)],~~~~2|b|\ge|c|.
\end{equation}
Integral (\ref{eintux}) can then be easily taken,
\begin{equation}                                        \label{ethuxc}
  \tg(cu)=\frac c{2b}\tg[b(x+x_0)],
\end{equation}
where an integration constant is inessential.
A solution of Eqs.~(\ref{eqthet}) and (\ref{eqphil}) has the form
\begin{equation*}
\begin{split}
  \cos\theta&=\sqrt{1-\frac{a^2}{c^2}}\sin[c(u+u_0)],~~~~|c|>|a|\ne0,
\\
  \tg(\vf+\vf_0)&=\frac ac\tg[c(u+u_0)].
\end{split}
\end{equation*}
An elementary geometrical analysis shows that solutions in a general situation
essentially coincide with degenerate solutions III and IV, but the rotation
axis rotates in the plane tilted at the angle $\arccos(a/c)$ to the $XY$ plane.

For static solutions, the energy of the two-dimensional $\MS\MO(3)$ model is
\begin{equation*}
  E=\int_0^L dx\left(\frac12\om^{\prime2}+(1-\cos\om)
  \left(\theta^{\prime2}+\sin^2\theta\vf^{\prime2}\right)\right).
\end{equation*}
Directly substituting the solutions in this integral, we obtain the values
\begin{equation*}
  E=\left\lbrace
\begin{aligned}
  0, &~~~~\text{degenerate solutions II},
\\
  \frac12b^2L, &~~~~\text{degenerate solutions I},
\\
  2b^2L,&~~~~\text{all other solutions}.
\end{aligned} \right.
\end{equation*}
The energy is finite for finite $L$. Degenerate solutions II correspond
to the absolute minimum because the energy is positive definite.
\section{Projective $\MR\MP^d$ models}
The Lagrangian for the Skyrme model written in form (\ref{elaomn})
can be naturally generalized to the case where the target space is
the real projective space $\MR\MP^d$ of arbitrary dimension. Let $k^i$,
$i=1,\dotsc,d$, be a unit vector $(k^2=1)$ in the Euclidean space $\MR^d$.
We consider a vector $k^i(x)$ and scalar $\om(x)$ fields on an arbitrary
Minkowski space, $x\in\MR^{1,n-1}$. Obviously, the Lagrangian
\begin{equation}                                        \label{elagep}
\begin{split}
  L=A\pl_\al\om^2+B(\pl_\al k\pl_\al k)
  +C(\pl_\al k\pl_\al k)\pl_\bt\om^2
  +D&(\pl_\al k\pl_\bt k)\pl_\al\om\pl_\bt\om
\\
  +E(\pl_\al k\pl_\al k)^2+F&(\pl_\al k\pl_\bt k)^2+G(\pl_\al\om^2)^2+U,
\end{split}
\end{equation}
where $A(\om),B(\om),C(\om),D(\om),E(\om),F(\om),G(\om)$, and $U(\om)$ are
arbitrary periodic (with period $2\pi$) functions of $\om$, is invariant
under transformations (\ref{einomn}). In the Lagrangian, we can introduce
an isotopic vector field $\om^i=\om k^i$ all of whose components are
independent, as in the Skyrme model. Because of equivalence relation
(\ref{eqrsog}), an isotopic vector field $\om^i$ can be considered to
take values inside a ball of radius $\pi$ with identified antipodal
points on the boundary sphere. This means that the field $\om^i$ takes
values in an arbitrary projective space $\MR\MP^d$.

This is the most general invariant Lagrangian depending only on first
derivatives of fields raised to a power not exceeding four. It is
invariant under the Poicar\'e group $\MI\MO(1,n-1)$ acting in the Minkowski
space $\MR^{1,n-1}$, the rotation group $\MO(d)$ acting in the target space
$\MR^d$, and translations (\ref{einomn}). The rotation group $\MO(d)$
can be extended to the group $\MO(d+1)$ (maximal symmetry group of the
projective space $\MR\MP^d$), but additional transformations act
nonlinearly on fields the $\om^i$.
\section{Conclusion}
We have considered the Skyrme model as a $\MS\MO(3)$ model where the target
space is the group manifold itself, and we used the explicit parameterization
of the group $\MS\MO(3)$ in terms of elements of its algebra. Because the
fundamental group of the rotation group is nontrivial, $\pi(\MS\MO(3))=\MZ_2$,
the model admits the existence of topologically nontrivial one-dimensional
solutions. The corresponding static solutions of the two-dimensional Skyrme
model are found and analyzed (degenerate solutions I in Sec.~7).

The form of the action for the $\MS\MU(2)$ and $\MS\MO(3)$ Skyrme models
is the same because their Lie algebras coincide. The difference
is in the equivalence relations in the parameter space. Hence, to
construct models in mathematical physics, we must not only specify the
action but also determine which manifold is the target space. This is
important because it may lead to topologically nontrivial solutions of the
Euler--Lagrange equations. For example, when considering spinors on a
three-dimensional Riemannian manifold, the $\MS\MO(3)$ connection enters
the covariant derivative, although the spinors transform under the
representation of unitary group $\MS\MU(2)$. This is because the group
$\MS\MO(3)$, not the group $\MS\MU(2)$, acts on an orthogonal vielbein
in the tangent space to a Riemannian manifold.

The $\MS\MO(3)$ model can be naturally generalized to the projective
space $\MR\MP^d$ of arbitrary dimension because the group $\MS\MO(3)$,
as a manifold, is diffeomorphic to the three-dimensional real projective
space $\MR\MP^3$. This allows defining a new class of projective
models depending on eight arbitrary periodic functions of one argument.
Topologically nontrivial one-dimensional solutions may exist in all
these models because $\pi(\MR\MP^d)=\MZ_2$ for $d\ge2$.

The $\MS\MO(3)$ model can be used in the geometric theory of defects
\cite{KatVol92,KatVol99}. The Riemann--Cartan geometry can be used to
describe the static distribution of dislocations and disclinations in
elastic media. The corresponding action is invariant under general
coordinate transformations and local $\MS\MO(3)$ rotations. The
equations of elasticity theory were recently used to fix the coordinate
system \cite{Katana03}. This allowed including the classical elasticity
theory in the geometric approach as a limiting case where defects are
absent. The $\MS\MO(3)$ model discussed in the present paper can be used
to fix local $\MS\MO(3)$ rotations because it is not invariant under these
rotations. In this case, the geometric theory of defects yields the
$\MS\MO(3)$ model in the absence of disclinations. For example, the
Lorenz gauge for the $\MS\MO(3)$ connection results in the equations of
the principal chiral field for the group $\MS\MO(3)$ in the absence of
disclinations. In the limit of no dislocations and disclinations, the
geometric defect theory can therefore be reduced to the equations of
the elasticity theory for the shift vector and to the equations of the
principal chiral field for the spin structure.

The author is very grateful to I.~V.~Volovich, R.~Dandoloff, E.~A.~Ivanov,
A.~I.~Pashnev, for the useful discussion and comments and to the Erwin
Schr\"odinger Institute (Vienna) for the hospitality.
This work was supported in part by the Russian Foundation for Basic
Research (grants 96-15-96131 and 02-01-01084).


\begin{thebibliography}{10}

\bibitem{Skyrme61}
T.~H.~R. Skyrme.
\newblock Nonlinear field theory.
\newblock {\em Proc.\ Roy.\ Soc. London}, A260:127--138, 1961.

\bibitem{ZaMaNoPi80E}
V.~E. Zakharov, S.~V. Manakov, S.~P. Novikov, and L.~P. Pitaevskii.
\newblock {\em The Inverse Scattering Method}.
\newblock Nauka, Moscow, 1980.
\newblock [in Russian]; English transl.: S.~P.~Novikov, S.~V.~Manakov,
  L.~P.~Pitaevskii, and V.~E.~Zakharov {\it The Inverse Scattering Method}
  Plenum, New York (1984).

\bibitem{Rajara82}
R.~Rajaraman.
\newblock {\em Solitons and Instantons in Quantum Field Theory}.
\newblock North--Holland, Amsterdam, 1982.

\bibitem{TakFad86E}
L.~A. Takhtadzhyan and L.~D. Faddeev.
\newblock {\em The Hamiltonian Methods in the Theory of Solitons}.
\newblock Nauka, Moscow, 1986.
\newblock [in Russian]; English transl.: L.~D.~Faddeev and L.~A.~Takhtajan {\it
  The Hamiltonian Methods in the Theory of solitons,} Berlin, Springer (1987).

\bibitem{Zakrze89}
W.~J. Zakrzewski.
\newblock {\em Low Dimensional Sigma Models}.
\newblock Adam Hilger, Bristol -- Philadelphia, 1989.

\bibitem{RybSan01E}
Yu.~P. Rybakov and V.~I. Sanyuk.
\newblock {\em Multidimensional Solitons}.
\newblock Russian Univ. of People's Friendship Publ., Moscow, 2001.
\newblock [in Russian].

\bibitem{BelPol75}
A.~A. Belavin and A.~M. Plyakov.
\newblock Metastable states of two-dimensional isotropic ferromagnet.
\newblock {\em JETP Letters}, 22(10):245--247, 1975.

\bibitem{EllDaw79}
J.~P. Elliott and P.~G. Dawber.
\newblock {\em Symmetry in Physics}, volume 2: Further Applications.
\newblock The Macmillan Press Ltd, London, 1979.

\bibitem{DuNoFo98E}
B.~A. Dubrovin, S.~P. Novikov, and A.~T. Fomenko.
\newblock {\em Modern geometry: Methods and Applications}.
\newblock Nauka, Moscow, fourth edition, 1998.
\newblock [In Russian]; English transl.\ prev.\ ed.: B.~A.~Dubrovin,
  A.~T.~Fomenko, and S.~P.~Novikov {\it Modern Geometry: Methods and
  Applications,} Part 1, {\it The Geometry of Surfaces, Transformation Groups,
  and Fields,} Springer, New York (1992).

\bibitem{Faddee77}
L.~D. Faddeev.
\newblock In search of multidimensional solitons.
\newblock In {\em Nonlocal. Nonlinear, and Nonrenormalizable Field Theories},
  pages 207--223, Dubna, 1977. Joint Inst.\ Nucl.\ Res.
\newblock [in Russian].

\bibitem{FadNie99A}
L.~Faddeev and A.~J. Niemi.
\newblock Partially dual variables in ${\MS\MU(2)}$ {Y}ang--{M}ills theory.
\newblock {\em Phys.\ Rev.\ Lett.}, 82:1624--1627, 1999.

\bibitem{FadNie97}
L.~Faddeev and A.~J. Niemi.
\newblock Knots and particles.
\newblock {\em Nature}, 387:58--66, 1997.

\bibitem{BatSut98}
R.~A. Battye and P.~Sutcliffe.
\newblock Knots as stable soliton solurions in a three-dimensional classical
  field theory.
\newblock {\em Phys.\ Rev.\ Lett.}, 81:4798--4801, 1998.

\bibitem{BatSut99}
R.~A. Battye and P.~Sutcliffe.
\newblock Solitons, links and knots.
\newblock {\em Proc.\ Roy.\ Soc.\ London}, A455:4305--4331, 1999.

\bibitem{HieSal99}
J.~Hietarinta and P.~Salo.
\newblock Faddeev--hopf knots: dynamics of linked unknots.
\newblock {\em Phys.\ Lett.}, B451:60--67, 1999.

\bibitem{KatVol92}
M.~O. Katanaev and I.~V. Volovich.
\newblock Theory of defects in solids and three-dimensional gravity.
\newblock {\em Ann.\ Phys.}, 216(1):1--28, 1992.

\bibitem{KatVol99}
M.~O. Katanaev and I.~V. Volovich.
\newblock Scattering on dislocations and cosmic strings in the geometric theory
  of defects.
\newblock {\em Ann.\ Phys.}, 271:203--232, 1999.

\bibitem{Katana03}
M.~O. Katanaev.
\newblock Wedge dislocation in the geometric theory of defects.
\newblock {\em Theor.\ Math.\ Phys.}, 135(2):733--744, 2003.

\end{thebibliography}
\end{document}